\documentclass[onecolumn,showpacs,amsmath,amssymb]{revtex4}
\usepackage{amsmath}
\usepackage{graphicx}
\usepackage{amsfonts}
\usepackage{dcolumn}
\usepackage{bm}
\usepackage{ulem}

\usepackage{CJK}

\begin{document}
\title{Nucleation of a three-state spin model on complex networks}

\author{Hanshuang Chen$^1$} \email{chenhshf@ahu.edu.cn}

\author{Chuansheng Shen$^{2}$}

\affiliation{ $^{1}$School of Physics and Material Science, Anhui
University, Hefei, 230039, People's Republic of China
\\$^2$Department of Physics, Anqing Normal University, Anqing, 246011,
China}

\date{\today}

\begin{abstract}
We study the metastability and nucleation of the Blume-Capel model
on complex networks, in which each node can take one of three
possible spin variables $\left\{ {-1, 0, 1} \right\}$. We consider
the external magnetic field $h$ to be positive, and let the chemical
potential $\lambda$ vary between $-h$ and $h$ in a low temperature,
such that the $1$ configuration is stable, and $-1$ configuration
and/or $0$ configuration are metastable. Combining the heterogeneous
mean-field theory with simulations, we show that there exist four
regions with distinct nucleation scenarios depending on the values
of $h$ and $\lambda$: the system undergoes a two-step nucleation
process from $-1$ configuration to $0$ configuration and then to $1$
configuration (region I); nucleation becomes a one-step process
without an intermediate metastable configuration directly from $-1$
configuration to $1$ configuration (region II(1)) or directly from
$0$ configuration to $1$ configuration (region II(2)) depending on
the sign of $\lambda$; the metastability of the system vanishes and
nucleation is thus irrelevant (region III). Furthermore, we show
that in the region I nucleation rates for each step intersect that
results in the occurrence of a maximum in the total nucleation rate.
\end{abstract}
 \maketitle

\section{Introduction}
Complex networks describe not only the pattern discovered
ubiquitously in the real world, but also provide a unified
theoretical framework to understand the inherent complexity in
nature \cite{Nature.393.440,Science.286.509}. A central topic in
this field is to unveil the relationship between the topology of a
network and dynamics taking place on it
\cite{RMP02000047,SIR03000167,PRP06000175, PRP08000093}. In
particular, phase transitions on complex networks have been a
subject of intense research in the field of statistical physics and
many other disciplines \cite{RMP08001275}. Extensive research
interests have focused on the onset of phase transitions in diverse
network topologies. Owing to the heterogeneity in degree
distribution, phase transitions on complex networks are drastically
different from those on regular lattices in Euclidean space. For
instance, degree heterogeneity can lead to a vanishing percolation
threshold \cite{PhysRevLett.85.4626}, the whole infection of disease
with any small spreading rate \cite{PRL01003200}, the Ising model to
be ordered at all temperatures
\cite{PHA02000260,PLA02000166,PRE02016104}, the disorder-order
transition in voter models \cite{EPL0768002}, synchronization to be
suppressed \cite{PhysRevLett.91.014101,PhysRevE.71.016116} and
different paths towards synchronization in oscillator network
\cite{PhysRevLett.98.034101}, spontaneous differentiation of
nonequilibrium pattern \cite{Nat10000544}, to list just a few.
However, there is much less attention paid to the dynamics of a
phase transition itself on complex networks, such as nucleation in a
first-order phase transition.

Nucleation is a fluctuation-driven process that initiates the decay
of a metastable state into a more stable one \cite{Kashchiev2000}.
Many important phenomena in nature, like crystallization
\cite{Nature011020}, glass formation \cite{PhysRevE.57.5707}, and
protein folding \cite{PNAS9510869}, are closely related to the
nucleation process. In the context of complex networks, the study of
the nucleation process is not only of theoretical importance for
understanding how a first-order phase transition happens in
networked systems, but also may have potential implications for
controlling fluctuation-driven system-wide transitions in real
situations, such as the transitions between different dynamical
attractors in neural networks \cite{PNAS04004341}, the genetic
switch between high- and low-expression states in gene regulatory
networks \cite{PNAS06008372,Plos09004872}, a new opinion
\cite{JSM0708026} or scientific paradigm formation
\cite{PhysRevLett.106.058701} as well as language replacement
\cite{CCP08000935,ACS0800357} in social networks, and spontaneous
traffic jamming \cite{epl2005}, synchronization
\cite{PhysRevLett.106.128701,PhysRevLett.108.168702}, cascading
failure \cite{Nature2010} and recovery \cite{NatPhys2014} close to
an explosive phase transition.

Recently, we have made a tentative step in the study of the
nucleation process of the two-state Ising model on complex networks,
where we have identified nucleation pathways, such as nucleating
from nodes with smaller degree on heterogeneous networks
\cite{PhysRevE.83.031110} and a multi-step nucleation process on
modular networks \cite{PhysRevE.83.046124}. In addition, a
size-effect of the nucleation rate on mean-field-type networks
\cite{PhysRevE.83.031110} and a nonmonotonic dependences of the
nucleation rate on the modularity of networks
\cite{PhysRevE.83.046124} and on the degree heterogeneity
\cite{JSTAT2013} were reported. However, many real systems possess
complicated free-energy landscape with several local minima where
phase transition happens usually via these intermediate metastable
states \cite{Wales2003}. The presence of intermediate metastable
states has been shown to play a key role in determining the pathway
and rate of nucleation. For example, it was recently reported that
an intermediate metastable phase can provide an easier pathway for
the growth of crystal nuclei from fluids, with implications for the
crystallization of protein and colloid
\cite{Science1997,PhysRevLett.96.046102,PhysRevLett.105.025701,PhysRevLett.107.175702,NatPhys2014(2)}.
Thus, it is natural to generalize the nucleation of the two-state
Ising model to a three-state spin model on complex networks in which
an intermediate metastable state may exist.

In this paper, we shall use the three-state Blume-Capel (BC) model
to investigate the nucleation on complex networks. The BC model is a
spin-1 Ising model that has been introduced, by Blume
\cite{Blume1966} and Capel \cite{Capel1966} independently, as a
model for magnetic systems and then applied to multicomponent fluids
\cite{PhysRevA.10.610}. The BC model defined on a two-dimensional
lattice has been previously used to study the metastability and
nucleation in the limit of zero temperature \cite{JSP1996} and in
the absence of external magnetic field \cite{PA2001}. Recently, a
reentrance phase transition has been observed in the BC model
defined on heterogeneous networks \cite{EPL2002}. Here, we show, by
using mean-field analysis and simulation, that there are four
distinct regions corresponding to different nucleation scenarios for
the networked BC model. Depending on the model's parameters, the
system undergoes either a two-step nucleation process with an
intermediate metastable state or a one-step nucleation process. We
also calculate the rates of nucleation by a rare-event sampling
method that agree with the theoretical predictions by evaluating the
free-energy barrier to nucleate.

\section{Model}
We consider the BC model defined on a network, where spin variable
of each node can take three possible values ${\sigma _i} \in \{  -
1,0,1\}$, and interacting according to the Hamiltonian
\begin{eqnarray}
\mathcal {H}= J\sum\limits_{i < j} {{a_{ij}}{{\left( {{\sigma _i} -
{\sigma _j}} \right)}^2} - \lambda \sum\limits_i {\sigma _i^2 -
h\sum\limits_i {{\sigma _i}} } },
\end{eqnarray}
where $J$ is the ferromagnetic interaction constant among nodes,
$\lambda$ and $h$ have the meaning of the chemical potential and the
external magnetic field imposed on each node, respectively. The
elements of the adjacency matrix of the network take $a_{ij} = 1$ if
nodes $i$ and $j$ are connected and $a_{ij} =0$ otherwise.

The present paper is devoted to the study of metastability and
nucleation of the networked three-state BC model at the low
temperature. For the purpose, we first consider the stability of the
system in the zero temperature limit. In this case, the (local)
stable equilibrium refer to the configurations with all the spins
equal to $-1$, $0$, $1$, respectively. For the sake of convenience,
we use $\uline{-1}$, $\uline{0}$, $\uline{1}$ to denote these stable
ordered configurations, respectively. Their energy are as follows:
$h-\lambda$, $0$, and $-h-\lambda$. Since we want to study the
nucleation from $\uline{-1}$ to $\uline{0}$, and then to
$\uline{1}$, we set $\uline{-1}<\uline{0}<\uline{1}$ as the relative
stabilities of these configurations. To the end, it is required that
$h-\lambda>0>-h-\lambda$, or equivalently, $-h<\lambda<h$ and $h>0$.
Due to the small thermal fluctuation at the low temperature, it is
expected that the behavior at the low temperature is similar to that
at the zero temperature. However, in the presence of small thermal
fluctuation the notations $\uline{-1}$, $\uline{0}$, $\uline{1}$
refer to the configurations with most instead of all the spins equal
to $-1$, $0$, $1$, respectively. Here, the temperature is fixed at
$T=5$ (in unit of $J/k_B$) throughout the paper where $k_B$ is the
Boltzmann constant.

\section{Theory and Simulation}
To proceed the heterogeneous mean-field theory, we first define
$X_k^{\left( \alpha \right)}$ as the probability that a node of
degree $k$ is in the state $\alpha \in \{-1,0,1\}$. The interaction
energy of an edge connecting a $k$-degree node and a $k'$-degree
node is thus written as,

\begin{eqnarray}
{E_{kk'}} & =& \nonumber  J\left[ {X_k^{\left( 1 \right)}\left(
{X_{k'}^{\left( 0 \right)} + 4X_{k'}^{\left( { - 1} \right)}}
\right) + X_k^{\left( 0 \right)}\left( {X_{k'}^{\left( 1 \right)} +
X_{k'}^{\left( { - 1} \right)}} \right) + X_k^{\left( { - 1}
\right)}\left( {X_{k'}^{\left( 0 \right)} + 4X_{k'}^{\left( 1
\right)}} \right)} \right] \\ {\kern 1pt} {\kern 1pt} {\kern 1pt}
{\kern 1pt} {\kern 1pt} {\kern 1pt} {\kern 1pt} {\kern 1pt} {\kern
1pt} {\kern 1pt} {\kern 1pt} {\kern 1pt} {\kern 1pt} {\kern 1pt}
{\kern 1pt} {\kern 1pt} {\kern 1pt} {\kern 1pt} {\kern 1pt} & =&
J\left( {{Q_k} + {Q_{k'}} - 2{M_k}{M_{k'}}} \right),
\end{eqnarray}
where
\begin{subequations}
\begin{align}
M_k = X_k^{\left( 1 \right)} - X_k^{\left( { - 1} \right)}\\
Q_k = X_k^{\left( 1 \right)} + X_k^{\left( { - 1} \right)}
\end{align}
\end{subequations}
are the average magnetization and the average squared magnetization
of a node of degree $k$, respectively.

Plus the single-node energy, the total energy of the system can be
expressed as
\begin{eqnarray}
E &=& \nonumber \frac{1}{2}N\sum\limits_k {P(k)k\sum\limits_{k'} {P(k'|k)} } {E_{kk'}} - \lambda N\sum\limits_k {P(k)} {Q_k} - hN\sum\limits_k {P(k)} {M_k}\\
{\kern 1pt} {\kern 1pt} {\kern 1pt} {\kern 1pt} {\kern 1pt} {\kern
1pt} {\kern 1pt} {\kern 1pt} & =& JN\left\langle k \right\rangle
\left( {Q' - {{M'}^2}} \right) - \lambda NQ - hNM
\end{eqnarray}
where $P(k)$ is the degree distribution, and ${P(k'|k)}$ is the
conditional probability that a node of degree $k$ links to a node of
degree $k'$. We use $P(k'|k) = {{k'P(k')} \mathord{\left/
 {\vphantom {{k'P(k')} {\left\langle k \right\rangle }}} \right.
 \kern-\nulldelimiterspace} {\left\langle k \right\rangle }}$ under
 the consideration of without degree correlation, where ${\left\langle k
\right\rangle }$ is the average degree. In Eq.(4), we have used the
definitions,
\begin{subequations}
\begin{align}
M &= \sum\limits_k {P(k){M_k}} \\
Q &= \sum\limits_k {P(k){Q_k}} \\
M' &= \sum\limits_k {\frac{{kP(k)}}{{\left\langle k \right\rangle
}}{M_k}} \\
Q'&= \sum\limits_k {\frac{{kP(k)}}{{\left\langle k \right\rangle
}}{Q_k}}
\end{align}
\end{subequations}
where $M$ and $Q$ are the average magnetization and the average
squared magnetization per node, respectively. $M'$ and $Q'$ are the
average magnetization and the average squared magnetization of a
randomly chosen nearest node, respectively.

Furthermore, let us define $S_k$ as the entropy of a node of degree
$k$, the total entropy of the system is
\begin{eqnarray}
S = N\sum\limits_k {P(k)} {S_k},
\end{eqnarray}
with
\begin{eqnarray}
{S_k} =  - {k_B}\sum\limits_\alpha  {X_k^{\left( \alpha \right)}\ln
X_k^{\left( \alpha  \right)}}.
\end{eqnarray}
Combining Eqs.(4) and (6), we can get the expression of free energy,
$F = E-TS$.

In order to get the extrema of the free energy, we use the Lagrange
function,
\begin{eqnarray}
\phi  =  - f + \sum\limits_k {P(k){\mu _k}} \left( {1 -
\sum\limits_\alpha  {X_k^{(\alpha )}} } \right),
\end{eqnarray}
where $f=F/N$ is the average free energy per node, and $\mu_k$ is
the Lagrange multiplier to maintain the normalization condition. By
minimizing of Eq.(8) with respect to $X_k^{(\alpha )}$, one has,
\begin{eqnarray}
X_k^{(\alpha )} = \frac{{\exp \left[ { - \beta \left(
{\frac{{\partial e}}{{\partial X_k^{(\alpha )}}}} \right)}
\right]}}{{\sum\limits_\alpha  {\exp \left[ { - \beta \left(
{\frac{{\partial e}}{{\partial X_k^{(\alpha )}}}} \right)} \right]}
}},
\end{eqnarray}
with
\begin{eqnarray}
\left\{ \begin{array}{l}
\frac{{\partial e}}{{\partial X_k^{(1)}}} = Jk\left( {1 - 2M'} \right) - \lambda  - h\\
\frac{{\partial e}}{{\partial X_k^{(0)}}} = 0\\
\frac{{\partial e}}{{\partial X_k^{( - 1)}}} = Jk\left( {1 + 2M'}
\right) - \lambda  + h
\end{array} \right.
\end{eqnarray}
where $\beta=1/(k_BT)$ and $e=E/N$ is the average energy per node.

Substituting Eq.(9) into Eq.(3), we get

\begin{eqnarray}
{M_k} = \frac{{2\sinh \left[ {\beta \left( {2JkM' + h} \right)}
\right]}}{{2\cosh \left[ {\beta \left( {2JkM' + h} \right)} \right]
+ \exp \left[ {\beta \left( {Jk - \lambda } \right)} \right]}} \\
{Q_k} = \frac{{2\cosh \left[ {\beta \left( {2JkM' + h} \right)}
\right]}}{{2\cosh \left[ {\beta \left( {2JkM' + h} \right)} \right]
+ \exp \left[ {\beta \left( {Jk - \lambda } \right)} \right]}}
\end{eqnarray}
Furthermore, inserting Eq.(11) into Eq.(5c), we get a
self-consistent equation of $M'$ that can be numerically solved,
\begin{eqnarray}
M' = \sum\limits_k {\frac{{kP(k)}}{{\left\langle k \right\rangle }}}
\frac{{2\sinh \left[ {\beta \left( {2JkM' + h} \right)}
\right]}}{{2\cosh \left[ {\beta \left( {2JkM' + h} \right)} \right]
+ \exp \left[ {\beta \left( {Jk - \lambda } \right)} \right]}}
\end{eqnarray}
Once the solutions of $M'$ are determined, all quantities will be
obtained, including the extrema of free energy.

\begin{figure}
\begin{center}
\includegraphics [width=10cm]{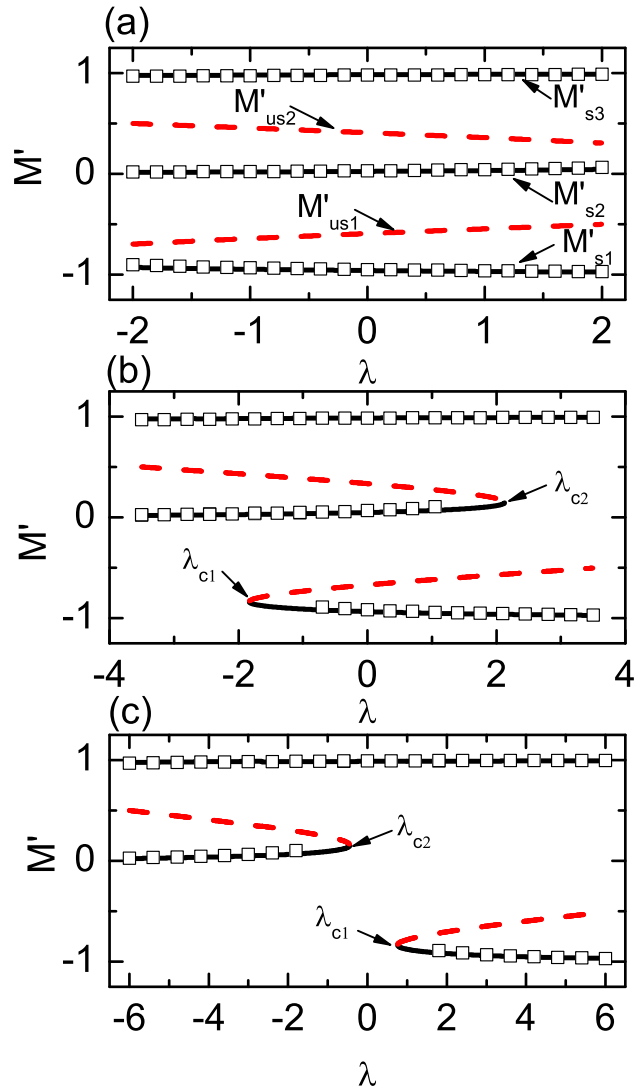}
\caption{(color online) The solutions of $M'$ as a function of
$\lambda$ at three different $h$: $h=2$ (a), $h=3.5$ (b), and $h=6$
(c). The solid and dashed lines depict the stable and unstable
solutions of $M'$, respectively, given by the analysis Eq.(15) with
a Poissonian degree distribution.  The square points are given by MC
simulations in $N=1000$ ER random networks. The other parameters are
$\left\langle k \right\rangle=20$ and $T=5$. \label{fig1}}
\end{center}
\end{figure}

To begin with, we consider a Erd\"os-R\'enyi (ER) random network
with the Poissonian degree distribution $P(k) = {{{{\left\langle k
\right\rangle }^k}{e^ - }^{\left\langle k \right\rangle }}
\mathord{\left/
 {\vphantom {{{{\left\langle k \right\rangle }^k}e^{-\left\langle k \right\rangle }} {k!}}} \right.
 \kern-\nulldelimiterspace} {k!}}$ \cite{ER1959}. The average degree we use is $\left\langle k
\right\rangle=20$. By numerically solving Eq.(13), we plot the
solutions of $M'$ as a function of $\lambda$ for three typical
values of $h$, as shown in Fig.1. The stable and unstable solutions
are depicted as the solid and dashed lines, respectively. When $h$
is small, for example $h=2$ in Fig.1(a), there are five solutions in
the whole allowable range of $\lambda  \in [ - h,h]$, in which three
of them are stable. Let $M'_{s_1}$, $M'_{s_2}$, and $M'_{s_3}$
denote these stable solutions around $-1$, $0$, and $1$,
respectively. These stable solutions give the three stable states
$\uline{-1}$, $\uline0$, and $\uline1$, respectively. The other two
unstable solutions, $M'_{us_1}$ and $M'_{us_2}$, give the two
transition states from $\uline{-1}$ to $\uline0$ and from $\uline0$
to $\uline1$, respectively. When $h$ becomes relatively larger, as
shown in Fig.1(b), the stable solution $M'_{s_1}$ and the unstable
$M'_{us_1}$ collide and annihilate each other at
$\lambda=\lambda_{c_1}$ via a a saddle-node bifurcation. Meanwhile,
$M'_{s_2}$ and the unstable $M'_{us_2}$ collide and vanish at
$\lambda=\lambda_{c_2}>\lambda_{c_1}$ in the same way. In this case,
the property of solutions can be classified into three regions
depending on the value of $\lambda$. For $\lambda<\lambda_{c_1}$,
there are two stable solutions, $M'_{s_2}$ and $M'_{s_3}$, and one
unstable solution $M'_{us_2}$. For $\lambda>\lambda_{c_2}$, there
are also two stable solutions and one unstable solution, but they
are $M'_{s_1}$ $M'_{s_3}$, and $M'_{us_1}$. While for
$\lambda_{c_1}<\lambda<\lambda_{c_2}$, the property of solutions is
the same as in Fig.1(a). With further increasing $h$, as shown in
Fig.1(c), $\lambda_{c_1}$ shifts to a larger value, and at the same
time $\lambda_{c_2}$ shifts to a smaller value, so that
$\lambda_{c_1}>\lambda_{c_2}$. In this case, there is a single
stable solution $M'_{s_3}$ at $\lambda_{c_2}<\lambda<\lambda_{c_1}$.

For comparison, we have performed Monte Carlo simulations in ER
random networks with network size $N=1000$ to compute the steady
state values of $M'$. The simulations start from numerous various
initial configurations to sample all possible steady state values of
$M'$, where $M'$ can be conveniently computed as $M'=
{{\sum\nolimits_{i = 1}^N {{k_i}{\sigma _i}} } \mathord{\left/
 {\vphantom {{\sum\nolimits_{i = 1}^N {{k_i}{\sigma _i}} } {(\left\langle k \right\rangle N)}}} \right.
 \kern-\nulldelimiterspace} {(\left\langle k \right\rangle N)}}$
with $k_i$ being the degree of node $i$. The simulation results are
added into Fig.1 as shown by square points. It is clear that the
simulation results are in excellent agreement with our theoretical
estimations.

To get a global view, we have plotted the phase diagram in the
$h\sim\lambda$ plane, as shown in Fig.2. The phase diagram is
divided into four different regions according to the property of
solutions of $M'$, separating by the lines of $\lambda_{c_1}\sim h$
and $\lambda_{c_2}\sim h$. In region I, $M'$ have five solutions:
three of them are stable (corresponding to three stable states
$\uline{-1}$, $\uline0$, and $\uline1$), and the others are unstable
(two transition states from $\uline{-1}$ to $\uline0$ then to
$\uline1$). In region II(1), $M'$ have three solutions: two of them
are stable, and the other is unstable (two stable states
$\uline{-1}$, and $\uline1$ and one transition state). In region
II(2), $M'$ have also three solutions, but the two stable states are
$\uline0$, and $\uline1$ and one transition state between them. In
region III, $M'$ has only one stable solution that corresponds to
the state $\uline1$. Also, we have given the simulation results of
$\lambda_{c_1}(h)$ and $\lambda_{c_2}(h)$, as depicted by square
points in Fig.2. One can see that there exist some mismatches
between the theory and simulation. This is because that near the
boundaries the lifetimes of metastable states are rather short (or
have very low free-energy barrier to nucleate that will be
illustrated later), so that such metastable states are hard to
identify in the simulations.

\begin{figure}
\begin{center}
\includegraphics [width=10cm]{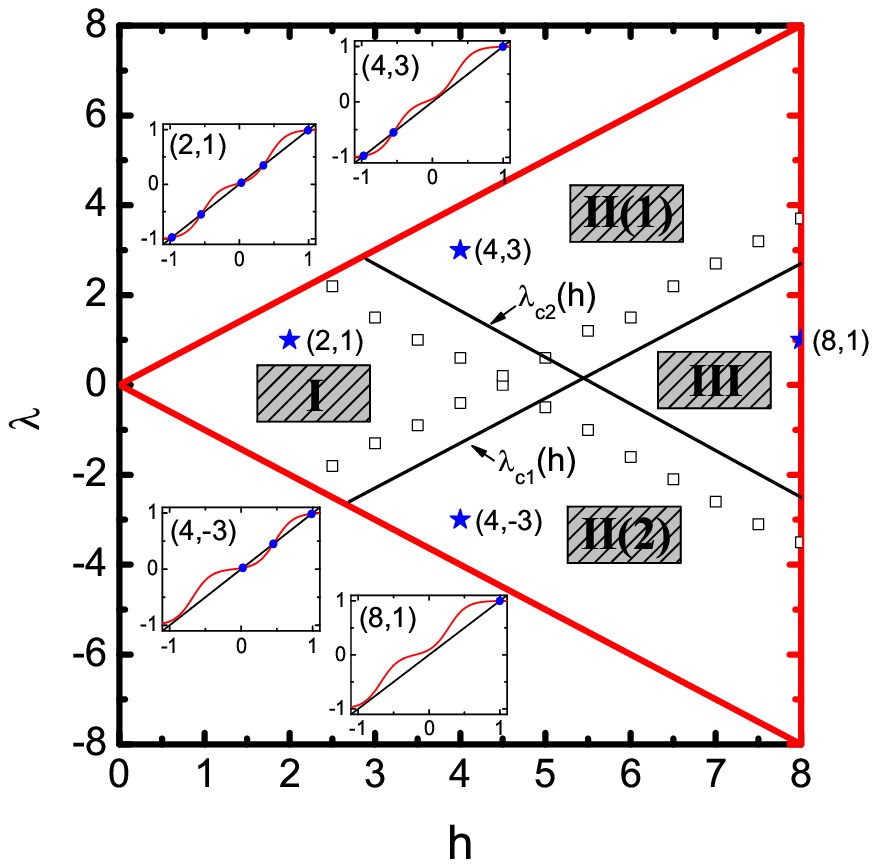}
\caption{(color online) Phase diagram in the $h\sim\lambda$ plane.
The phase diagram is divided into four different regions according
to the property of solutions of $M'$, separating by the lines of
$\lambda_{c_1}\sim h$ and $\lambda_{c_2}\sim h$. In region I, $M'$
have five solutions: three of them are stable (corresponding to
three stable states $\uline{-1}$, $\uline0$, and $\uline1$), and the
others are unstable (two transition states from $\uline{-1}$ to
$\uline0$ then to $\uline1$). In region II(1), $M'$ have three
solutions: two of them are stable, and the other is unstable (two
stable states $\uline{-1}$, and $\uline1$ and one transition state).
In region II(2), $M'$ have also three solutions, but the two stable
states are $\uline0$, and $\uline1$ and one transition state between
them. In region III, $M'$ have only one stable solution, and the
corresponding state is $\uline1$. Inset: Graphic solution of Eq.(15)
for four typical $(h, \lambda)$ points (stars in phase diagram)
chosen in four different regions. Square points give the simulation
results of $\lambda_{c_1}(h)$ and $\lambda_{c_2}(h)$. The other
parameters are the same as those in Fig.1. \label{fig2}}
\end{center}
\end{figure}

We have also performed the calculations in Barab\'asi-Albert (BA)
scale-free networks \cite{Science.286.509} with the same size and
average degree, and found that the phase diagram is almost the same
as that in ER random network (not shown here). That is, the phase
diagram is not almost affected by network topology.

In Fig.3, we have schematically demonstrated the nucleation process
in different regions. In region I, the system undergoes a two-step
nucleation process. The first stage is the nucleating of the
metastale $\uline 0$ from $\uline -1$. Subsequently, in the second
stage, the transition from $\uline 0$ to $\uline 1$ happens via the
nucleating of $\uline 1$. The free-energy barrier of the two-step
nucleation are $\Delta {F_{\uline{-1} \to \uline{0}}} = N\Delta
{f_{\uline{-1} \to \uline{0}}} = N(f_{M'_{US1}} - f_{M'_{S1}})$ and
$\Delta {F_{\uline{0} \to \uline{1}}} = N\Delta {f_{\uline{0} \to
\uline{1}}} = N(f_{M'_{US2}} - f_{M'_{S2}})$, respectively. Herein,
we use the notation $f_{M'_{*}}$ to denote free energy per node at
$M'=M'_{*}$. In regions II(1), since the state $\uline{0}$ is no
longer present, nucleation happens via a one-step process directly
from $\uline{-1}$ to $\uline{1}$. The resulting free-energy barrier
is $\Delta {F_{\uline{-1} \to \uline{1}}} = N\Delta {f_{\uline{-1}
\to \uline{1}}} = N(f_{M'_{US1}} - f_{M'_{S1}})$. In region II(2),
since the state $\uline{-1}$ ceases to exist, nucleation also
proceeds by one-step process from $\uline{0}$ to $\uline{1}$, and
the corresponding free-energy barrier is $\Delta {F_{\uline{0} \to
\uline{1}}}$. In region III, the only stable state is $\uline{1}$
and the nucleation is thus irrelevant.

\begin{figure}
\begin{center}
\includegraphics [width=10cm]{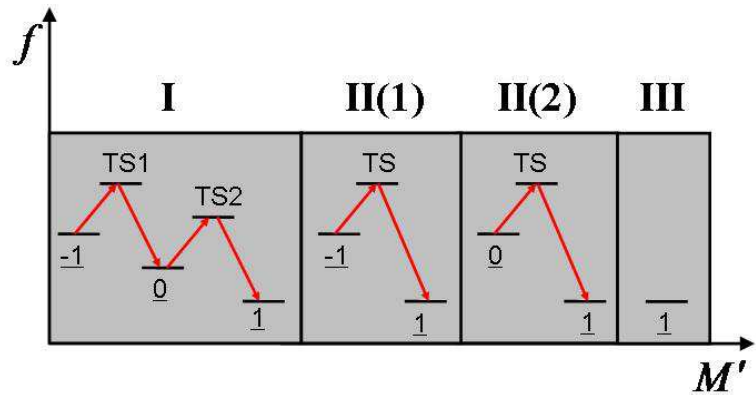}
\caption{(color online) Schematic plot for nucleation and growth in
different regions. In region I, nucleation from $\uline {-1}$ to
$\uline1$ proceeds in a two-step process via an intermediate
metastable state $\uline 0$. In both regions II(1) and II(2),
nucleation proceeds in a one-step process. In region III, the only
state $\uline 1$ is stable and thus nucleation is irrelevant.
\label{fig3}}
\end{center}
\end{figure}

\begin{figure}
\begin{center}
\includegraphics [width=12cm]{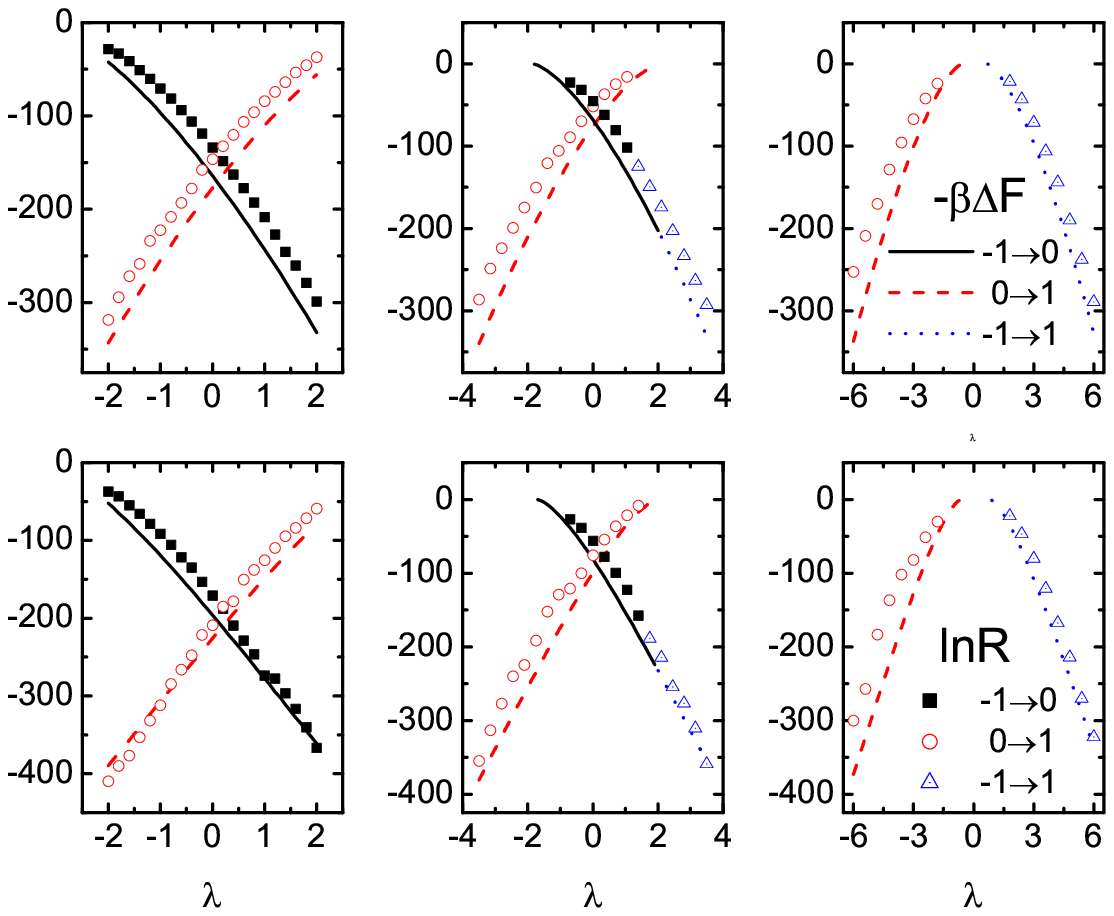}
\caption{(color online) The free energy barrier $-\beta \Delta F$
and the logarithm of rate of nucleation $\ln R$, obtained from the
HMF analysis (lines) and FFS simulations (symbols), as a function of
$\lambda$ at three different $h$: $h=2$, $3.5$, $6$ (from left to
right). Top and bottom panels show the results in ER networks and BA
networks, respectively. The other parameters are the same as those
in Fig.1. \label{fig4}}
\end{center}
\end{figure}

Since nucleation rate $R$ is exponentially dependent on $-\beta
\Delta F$, $R \sim \exp(-\beta \Delta F)$, at the top and bottom
panels of Fig.4 we show that $-\beta \Delta F$ as a function of
$\lambda$ at three typical different $h$: $h=2$, $3.5$, $6$ (from
left to right) in ER networks (top panels) and BA networks (bottom
panels), respectively (shown by lines). To validate the analytical
results, computer simulation for calculating nucleation rate is
desirable.

However, nucleation is an activated process that occurs extremely
slow, and brute-force simulation for observing nucleation process is
thus prohibitively expensive. To overcome this difficulty, we will
employ a recently developed simulation method, forward flux sampling
(FFS) \cite{PRL05018104,JCP07114109}. This method allows us to
calculate nucleation rate and determine the properties of ensemble
toward nucleation pathways. This method uses a series of interfaces
in phase space between the initial and final states to force the
system from the initial state $A$ to the final state $B$ in a
ratchet-like manner. Before the simulation begins, an order
parameter $r$ is first defined, such that the system is in state $A$
if $r<r_0$ and it is in state $B$ if $r>r_n$. A series of
nonintersecting interfaces $r_i$ ($0<i<n$) lie between states $A$
and $B$, such that any path from $A$ to $B$ must cross each
interface without reaching $r_{i+1}$ before $r_i$. The algorithm
first runs a long-time simulation which gives an estimate of the
flux $\bar \Phi_{A,0}$ escaping from the basin of $A$ and generates
a collection of configurations corresponding to crossings of
interface $r_0$. The next step is to choose a configuration from
this collection at random and use it to initiate a trial run which
is continued until it either reaches $r_1$ or returns to $r_0$. If
$r_1$ is reached, store the configuration of the end point of the
trial run. Repeat this step, each time choosing a random starting
configuration from the collection at $r_0$. The fraction of
successful trial runs gives an estimate of of the probability of
reaching $r_1$ without going back into $A$, $P\left( {r_1 |r_0}
\right)$. This process is repeated, step by step, until $r_n$ is
reached, giving the probabilities $P\left( {r_{i+1} |r_i} \right)$
($i=1, \cdots,n-1$). Finally, we get the nucleation rate $R$ from
$A$ to $B$ as
\begin{equation}
R=\bar \Phi_{A,0} P\left( {r_{n} |r_0} \right) =\bar \Phi_{A,0}
\prod\nolimits_{i=0}^{n-1}{P\left( {r_{i + 1} |r_i} \right)},
\label{eq2}
\end{equation}
where $P\left( {r_{n} |r_0} \right)$ is the probability that a
trajectory crossing $r_0$ in the direction of $B$ will eventually
reach $B$ before returning to $A$.

For comparison, we have also added the simulation results to Fig.4
(shown by symbols). On one hand, the analytical results are in well
agreement with the simulation ones. On the other hand, the results
in ER random networks and in BA scale-free networks are
qualitatively the same.

For $h=2$, nucleation is a two-step process in the whole allowable
range of $\lambda$, and the corresponding nucleation rates are
$R_{\uline{-1}\to\uline{0}}$ and $R_{\uline{0}\to\uline{1}}$,
respectively. As $\lambda$ increases, $R_{\uline{-1}\to\uline{0}}$
decreases monotonically, but $R_{\uline{0}\to\uline{1}}$ increases
monotonically. In this case, the total nucleation rate $R$ is
expressed as
$R=(R_{\uline{-1}\to\uline{0}}^{-1}+R_{\uline{0}\to\uline{1}}^{-1})^{-1}$.
It can be seen that $R$ is dominantly determined by the smaller of
$R_{\uline{-1}\to\uline{0}}$ and $R_{\uline{0}\to\uline{1}}$.
Therefore, there exists a maximal $R$ at $\lambda = \lambda_{opt}$
where $R_{\uline{-1}\to\uline{0}}$ and $R_{\uline{0}\to\uline{1}}$
intersect. Here, $\lambda_{opt}\simeq0.1$ for ER random networks and
$\lambda_{opt}\simeq0.2$ for BA scale-free networks that are robust
to different $h$. For $h=3.5$, nucleation is a two-step process in a
certain range of $\lambda$ around zero, while in other range
nucleation becomes a one-step process. The resulting total
nucleation rate has also a maximum at $\lambda = \lambda_{opt}$. For
$h=6$, nucleation is a one-step process and the corresponding
nucleation rate increases as $\lambda$ gradually approaches zero
until nucleation becomes irrelevant when $\lambda$ crosses the line
of $\lambda_{c_1}(h)$ or $\lambda_{c_2}(h)$.

\section{Conclusion}
In conclusion, we have studied the nucleation of the three-state BC
model on complex networks. By using the heterogeneous mean-field
theory and simulations, we have found that there exist four distinct
regions with different nucleation scenarios depending on the model's
parameters: the external field $h$ and the chemical potential
$\lambda$. For a small $h$, or for a moderate $h$ and simultaneously
a small $\left| \lambda \right|$, the system goes through a two-step
nucleation process from $\uline{-1}$ configuration to $\uline{0}$
configuration and then to $\uline{1}$ configuration. For a moderate
$h$ or a large $h$ and simultaneously a large $\left| \lambda
\right|$, nucleation is a one-step process without an intermediate
metastable configuration directly from $\uline{-1}$ configuration to
$\uline{1}$ configuration or directly from $\uline{0}$ configuration
to $\uline{1}$ configuration depending on the sign of $\lambda$. For
a large $h$ and simultaneously a small $\left| \lambda \right|$, the
metastability of the system vanishes and nucleation is thus
irrelevant. Moreover, we have calculated the nucleation rates and
found that in the two-step nucleation region there exists a maximum
for the total nucleation rate. All the results are demonstrated in
ER random networks and in BA scale-free networks and are
qualitatively the same for different network topologies.
Quantitatively, the optimal $\lambda_{opt}$ at which the total
nucleation rate is maximal in ER random networks is less than that
in more heterogeneous BA scale-free networks.

\begin{acknowledgments}
We acknowledge supports from the National Science Foundation of
China (11205002, 11475003, 61473001), ¡°211 project¡± of Anhui
University (02303319-33190133), and Anhui Provincial Natural Science
Foundation (1408085MA09).
\end{acknowledgments}


\end{document}